\documentclass[a4paper]{jpconf}
\usepackage{graphicx}
\usepackage{amssymb}
\usepackage{textcomp}
\begin{document}
\title{Muon Hunter: a Zooniverse project}

\author{R.~Bird$^{1}$, M.~K.~Daniel$^{2}$, H.~Dickinson$^{3}$, Q.~Feng$^{4,7}$, 
L.~Fortson$^{3}$, A.~Furniss$^{5}$, J.~Jarvis$^{6}$, R.~Mukherjee$^{7}$, R.~Ong$^{1}$, 
I.~Sadeh$^{8}$ and D.~Williams$^{9}$.}

\address{$^{1}$University of California, Los Angeles, 
Department of Physics and Astronomy, Physics and Astronomy Building, 430 Portola Plaza, Box 951547, Los Angeles, CA~90095.USA.} 
\address{$^{2}$Whipple Observatory, 670 Mt Hopkins Road, Amado, AZ~85645. USA.}
\address{$^{3}$University of Minnesota, School of Physics and Astronomy, 116 Church St. SE, Minneapolis, MN~55455. USA.}
\address{$^{4}$Physics Department, McGill University, 3600 University Street, Montreal, QC H3A~2T8. Canada.}
\address{$^{5}$California State University East Bay, 25800 Carlos Bee Blvd, Hayward, CA~94542. USA.}
\address{$^{6}$ASTERICS, DECS Workpackage, School of Physical Sciences, Robert Hooke Building, Open University, Walton Hall, Milton Keynes, Buckinghamshire, MK7~6AA. UK.}
\address{$^{7}$Columbia University, Barnard College, Deparment of Physics and Astronomy, 3009~Broadway, New York, NY~10027. USA.}
\address{$^{8}$Deutsches Elektronen-Synchrotron (DESY), Platanenallee 6, 15738~Zeuthen, Germany.}
\address{$^{9}$SCIPP, Room 337 Natural Sciences 2, University of California, 1156 High Street, Santa Cruz, CA~95064. USA.}

\ead{michael.daniel@cfa.harvard.edu}

\begin{abstract}
The large datasets and often low signal-to-noise inherent to the raw data of modern astroparticle experiments calls out for increasingly sophisticated event classification techniques. Machine learning algorithms, such as neural networks, have the potential to outperform traditional analysis methods, but come with the major challenge of identifying reliably classified training samples from real data. Citizen science represents an effective approach to sort through the large datasets efficiently and meet this challenge. Muon Hunter is a project hosted on the Zooniverse platform, wherein volunteers sort through pictures of data from the VERITAS cameras to identify muon ring images. Each image is classified multiple times to produce a “clean” dataset used to train and validate a convolutional neural network model both able to reject background events and identify suitable calibration data to monitor the telescope performance as a function of time.
\end{abstract}

\section{Introduction}
The Very Energetic Radiation Imaging Telescope Array System (VERITAS) is composed of four 12\,m diameter imaging atmospheric Cherenkov telescopes (IACTs) sited at the Fred Lawrence Whipple Observatory(FLWO) in southern Arizona (USA) \cite{VERITAS}. Each dish is a tessellated reflector made up of 350 individual mirror facets in a Davis-Cotton configuration focused onto a 499-pixel photomultiplier tube camera for a total field of view of $3.5^\circ$. 
The primary field of research of VERITAS is in the ground based detection of very high energy (VHE, E$\gtrsim$100\,GeV) gamma-rays under a seemingly overwhelming 1000:1 background of cosmic-rays. In both cases the primary very energetic particle will cause a particle cascade in the atmosphere known as an extensive air shower. As the air shower particles move through the atmosphere they in turn generate Cherenkov light which can be picked up and imaged by the telescopes on the ground. The cosmic ray background is reduced to approximately 1:1 through the stereo imaging technique, where cosmic rays will tend to have larger, blotchier, more variable light distribution patterns to the smooth, elliptical images expected in the camera for gamma rays. The cosmic ray showers will also result in a large number of muon particles being generated, muons that pass close by the optical axis of the telescope will cause ring like images in the camera. As the muon path moves further from the centre of the camera, or as the direction of travel moves off from the optical axis, the images will only be partial rings, which can readily mimic the elliptical gamma-ray primary images, causing false positive classification. Muon rings, however, can also be useful: the amount of light expected in the ring is a well known quantity and so they can be used to provide an absolute calibration of the camera digitisation. It is for these two reasons -- background event rejection and calibration event identification -- that we are interested in being able to pick out the muon ring images from the non-muon cosmic ray and gamma ray ones.

Classification is a common task in experimental physics and with inherently large datasets it is easy to see why the use of machine learning algorithms has become increasingly popular. With a trigger rate of O(400\,Hz) and over 10,000 hours of observational data for the VERITAS telescopes it is not possible for any one person to sift through all of the data for events of interest. One powerful machine learning algorithm, convolutional neural networks (CNN), was used on a small batch of VHE gamma-ray data to detect and characterize muon events \cite{CNN}.
For developing supervised machine learning algorithms like the CNN, it is essential to assign correct labels to a large training dataset. 
In this work, we describe Muon Hunter\footnote{\texttt{www.muonhunter.org}}, a citizen-science project where volunteers label and parameterize muon and non-muon images in VHE gamma-ray data to provide high purity machine learning training samples.

\section{Muon Hunter}
The Muon Hunter experiment was developed in collaboration with the ASTERICS Horizon2020 project\footnote{\texttt{www.asterics2020.eu}}, and the classification interface is hosted by the Zooniverse\footnote{\texttt{www.zooniverse.org}} platform, where researchers in many disciplines can easily design and deploy a citizen-science project through the Zooniverse Project Builder\footnote{\texttt{www.zooniverse.org/labs}}. The Zooniverse is the world's largest and most powerful platform for people-powered research. In the ten years since the launch of the first Zooniverse project, Galaxy Zoo~\cite{GalaxyZoo}, the platform has expanded to over 100 projects in a wide variety of research areas from the physical to the medical and social sciences, and liberal arts. Via the Zooniverse anyone can be a researcher; you do not need any specialised background, training, or expertise to participate (though training in the project task is given on the website). One further noteworthy aspect of the Zooniverse is that it is not simply an outreach tool, volunteers and professionals make real discoveries together -- many of which result in published research papers and open source data sets. The Zooniverse projects are constructed with the specific aim of converting volunteers' efforts into measurable results and scientifically significant discoveries \cite{GalaxyZoo, PlanetHunters, PhysicsToday, PlanetFour}.

\begin{figure}
\begin{center}
\includegraphics[width=0.66\textwidth]{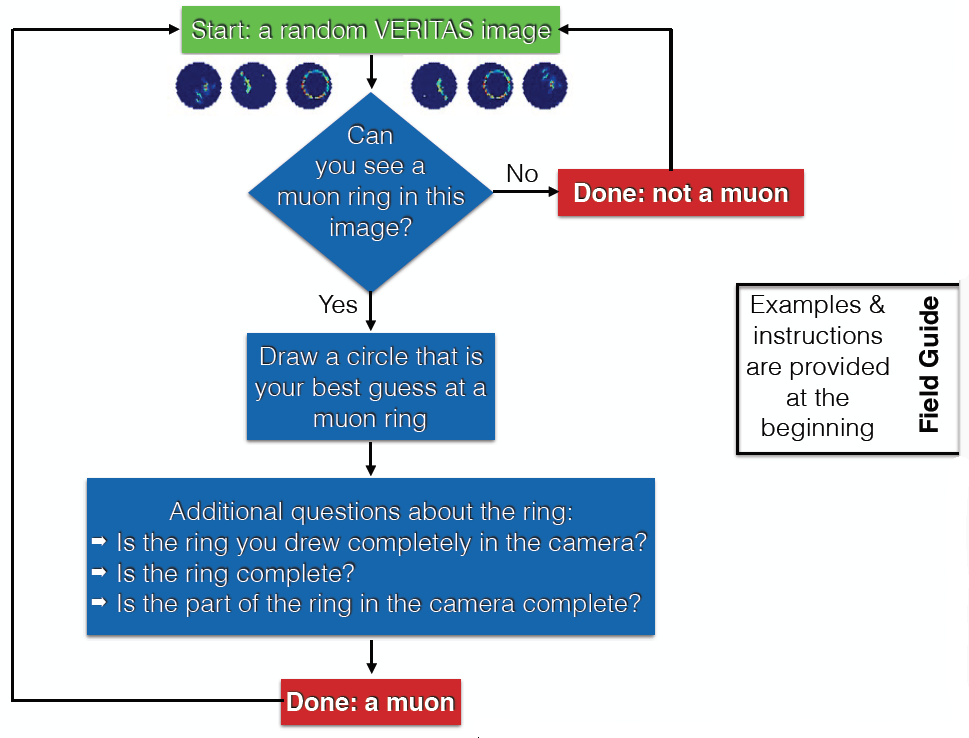}
\end{center}
\caption{\label{fig:WorkFlow}The workflow for classifying images as muons or non-muons.}
\end{figure}

\begin{figure}
\begin{center}
\includegraphics[width=0.25\textwidth]{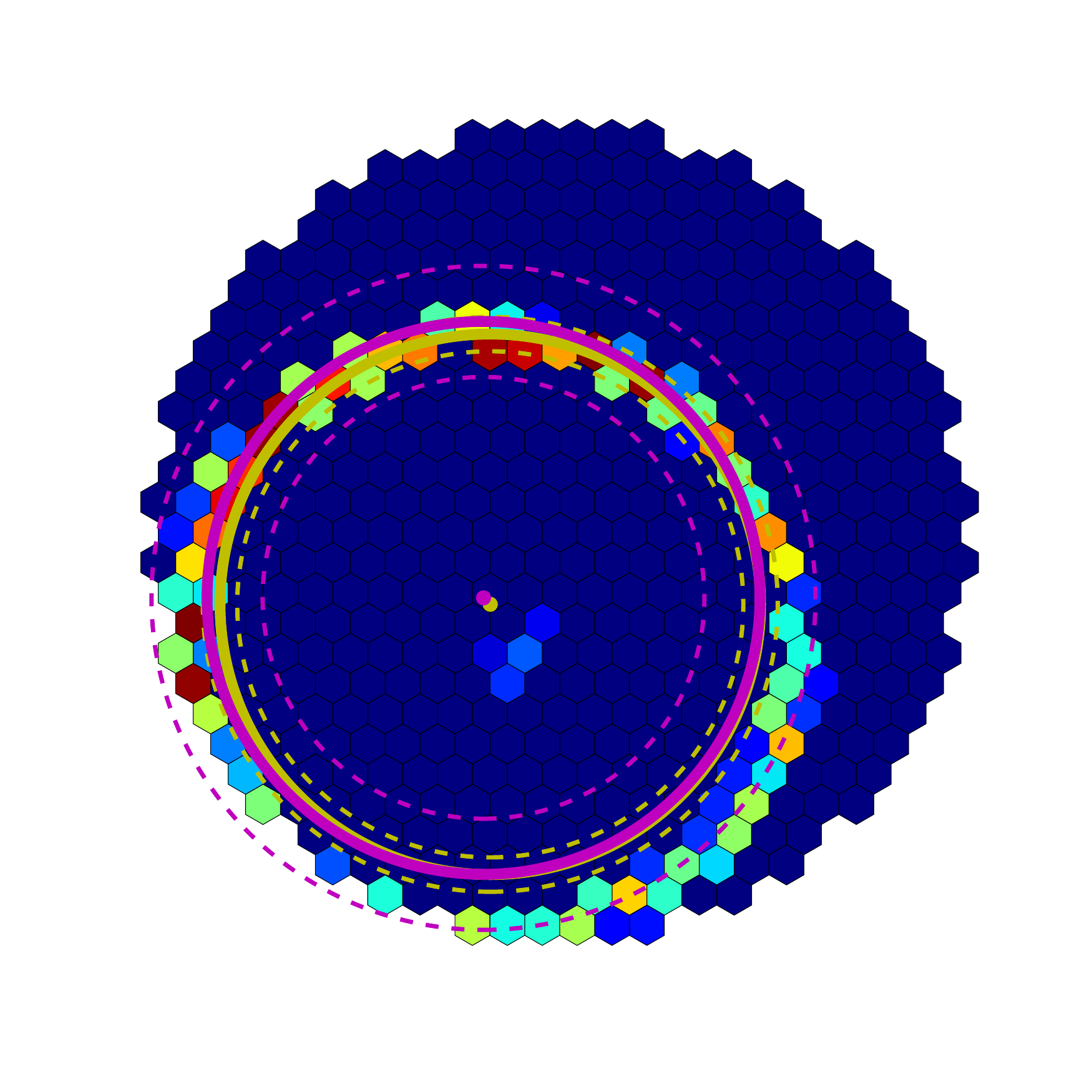}
\includegraphics[width=0.25\textwidth]{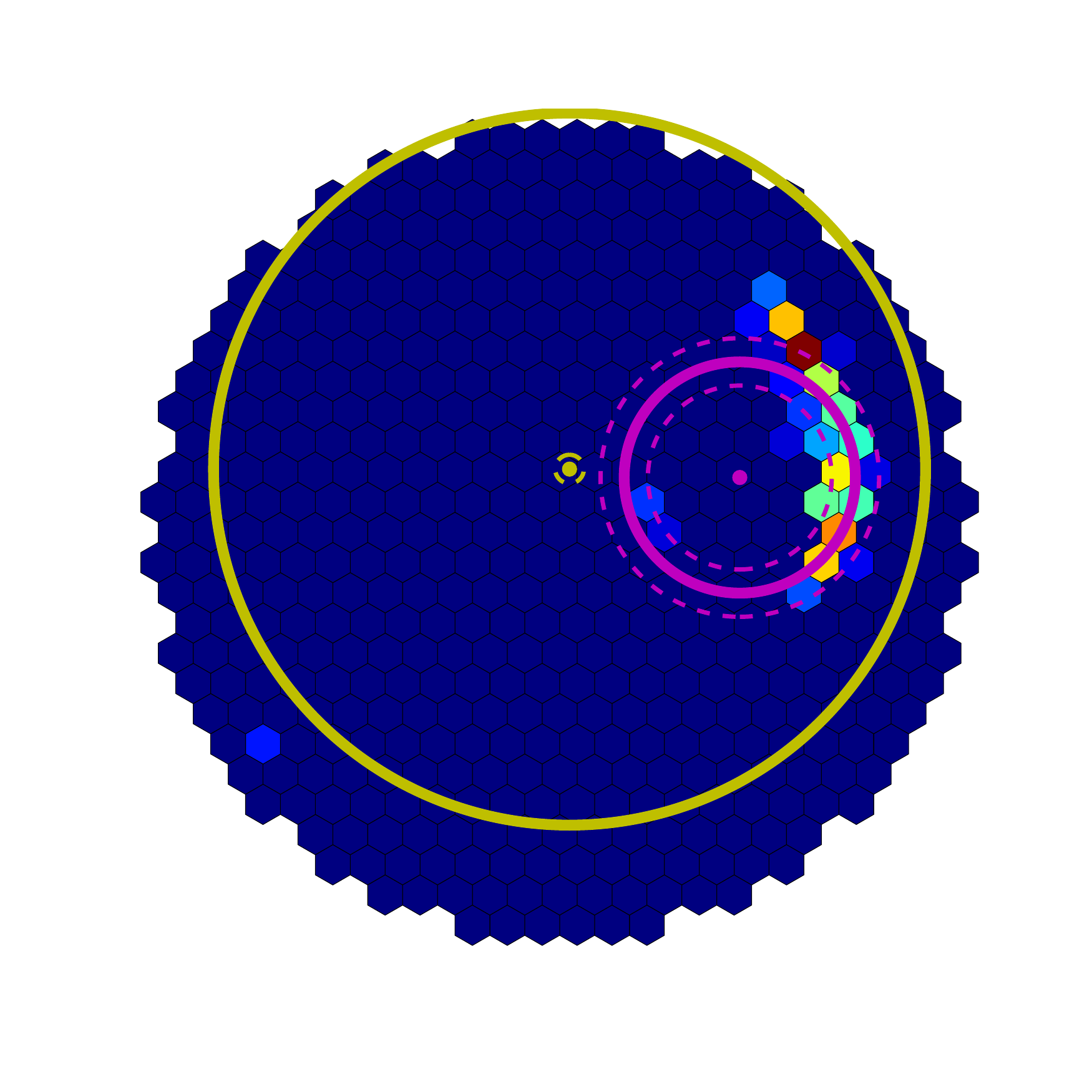}
\includegraphics[width=0.25\textwidth]{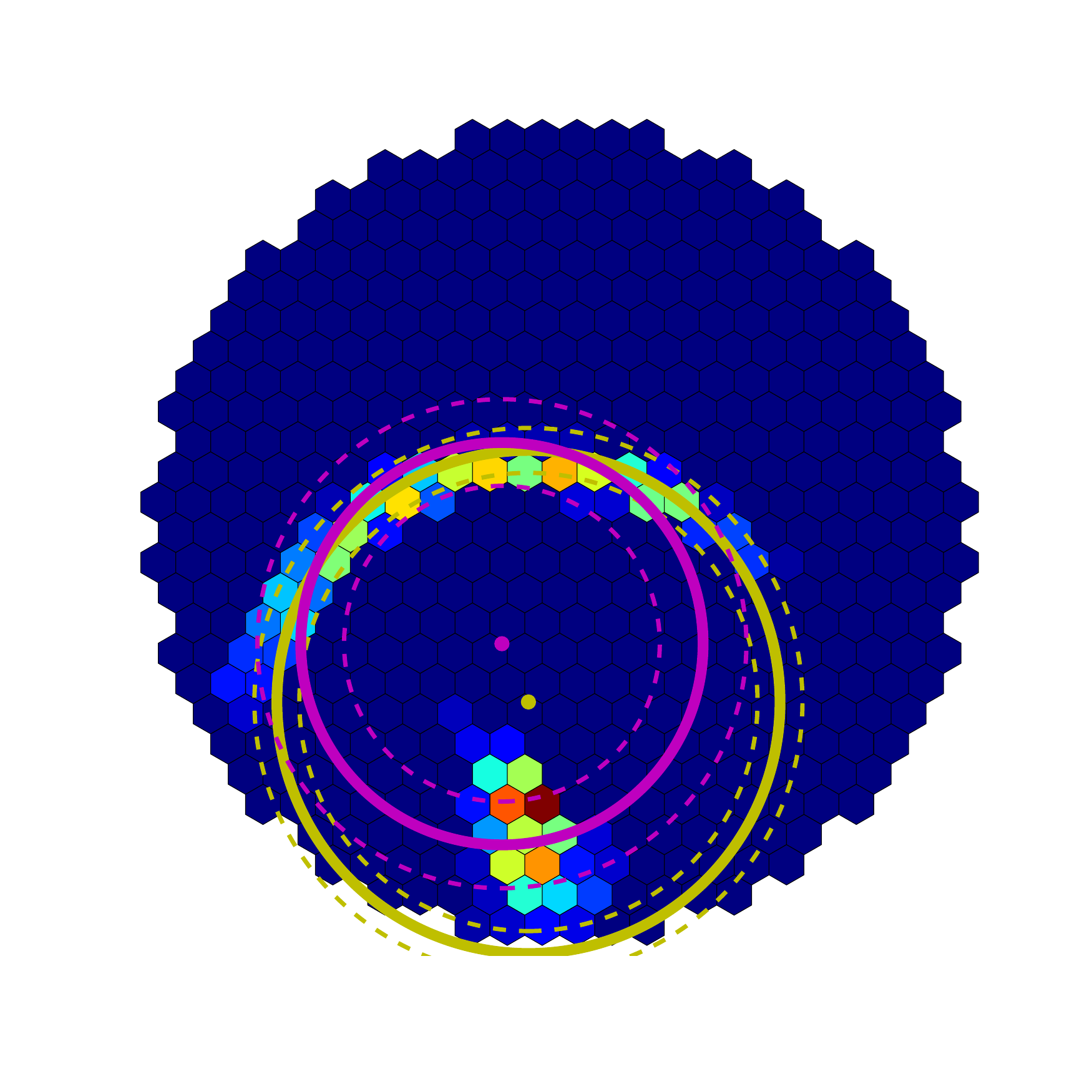}
\end{center}
\caption{\label{fig:MuonImages}Example muon ring images. Standard analysis results are shown in magenta and Muon Hunter volunteer input in yellow.}
\end{figure}

The key components of a Zooniverse project are data and workflows. The workflows are a sequence of specific tasks tailored by the researchers to gain insights on the data. The workflows are streamlined and optimised to balance quality of experience for the volunteers with the need to obtain accurate results. 
The workflow for identifying the presence of rings in an image is illustrated in Figure~\ref{fig:WorkFlow}. To help new users, a short tutorial, a mini course, and a detailed `About' page were available. Images were retired once a total of 15 volunteers had examined it and finished the workflow. A total of 137,515 VERITAS single-telescope images were served on the Muon Hunter website, with examples given in Figure~\ref{fig:MuonImages}. Most images are preprocessed using the standard VERITAS two-level cleaning \cite{VEGAS} based on the signal-to-noise ratio of each pixel, but a subset of uncleaned images were also uploaded to explore the effect of image cleaning on the results of classification. From 5,734 volunteers we received a total of $\sim$2.1~million classifications, half within the first week after the official launch of the project.

It was possible to draw multiple rings on one image, as sometimes more than one muon was recorded in an image. But this also allowed room for human error, especially when a user was unfamiliar with the workflow. After a user had completed the workflow of a subject image there was an option to discuss it in the Talk board. The Talk board enables interactions between experts and volunteers regarding specific images. Collections of interesting images are also established by users, including double muon rings and composite images with a muon ring and a background air shower, or particularly noteworthy, unexpected or unusual images that may be worthy of further investigation later.

\begin{figure}
\begin{center}
\includegraphics[width=0.45\textwidth]{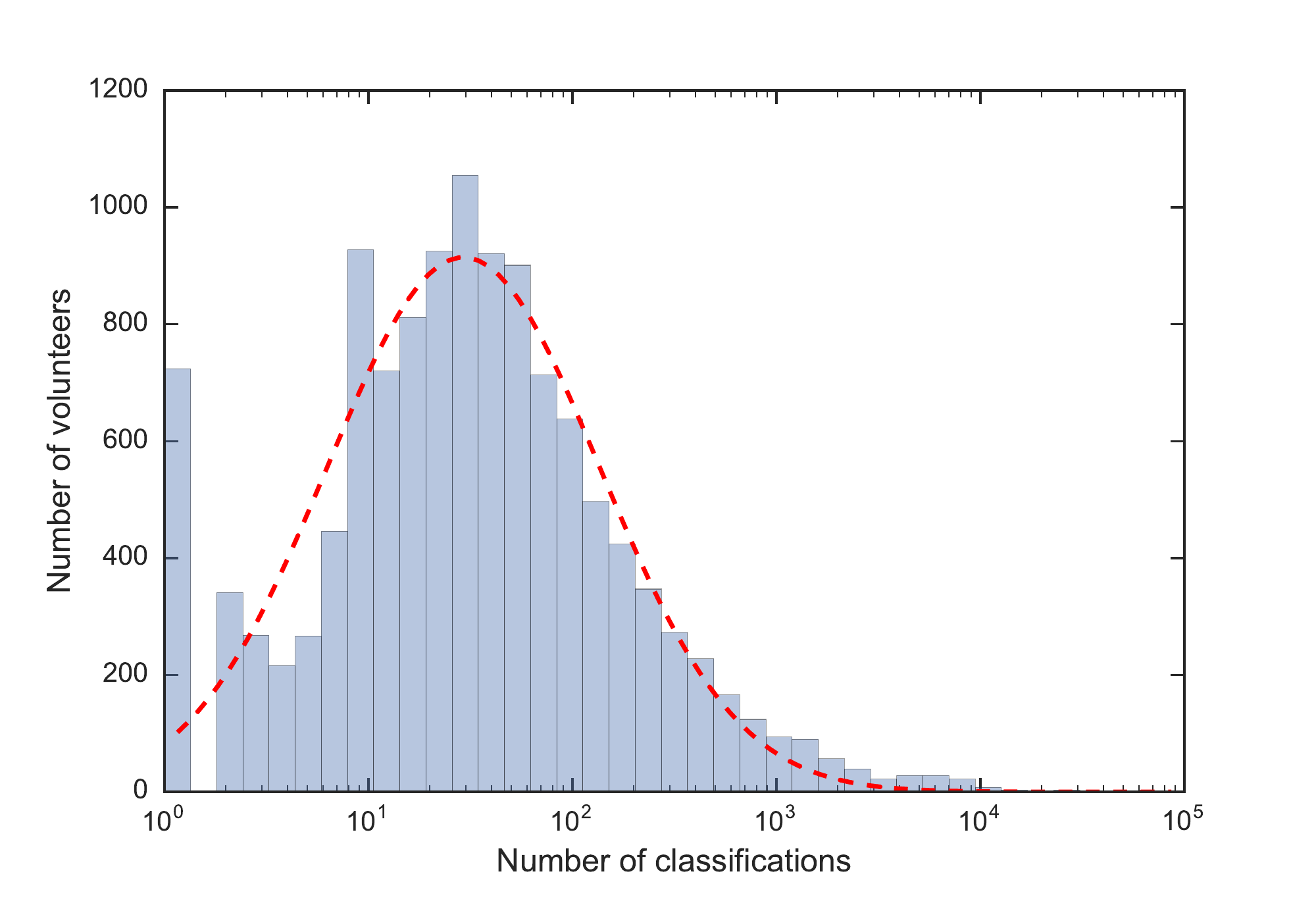}
\includegraphics[width=0.45\textwidth]{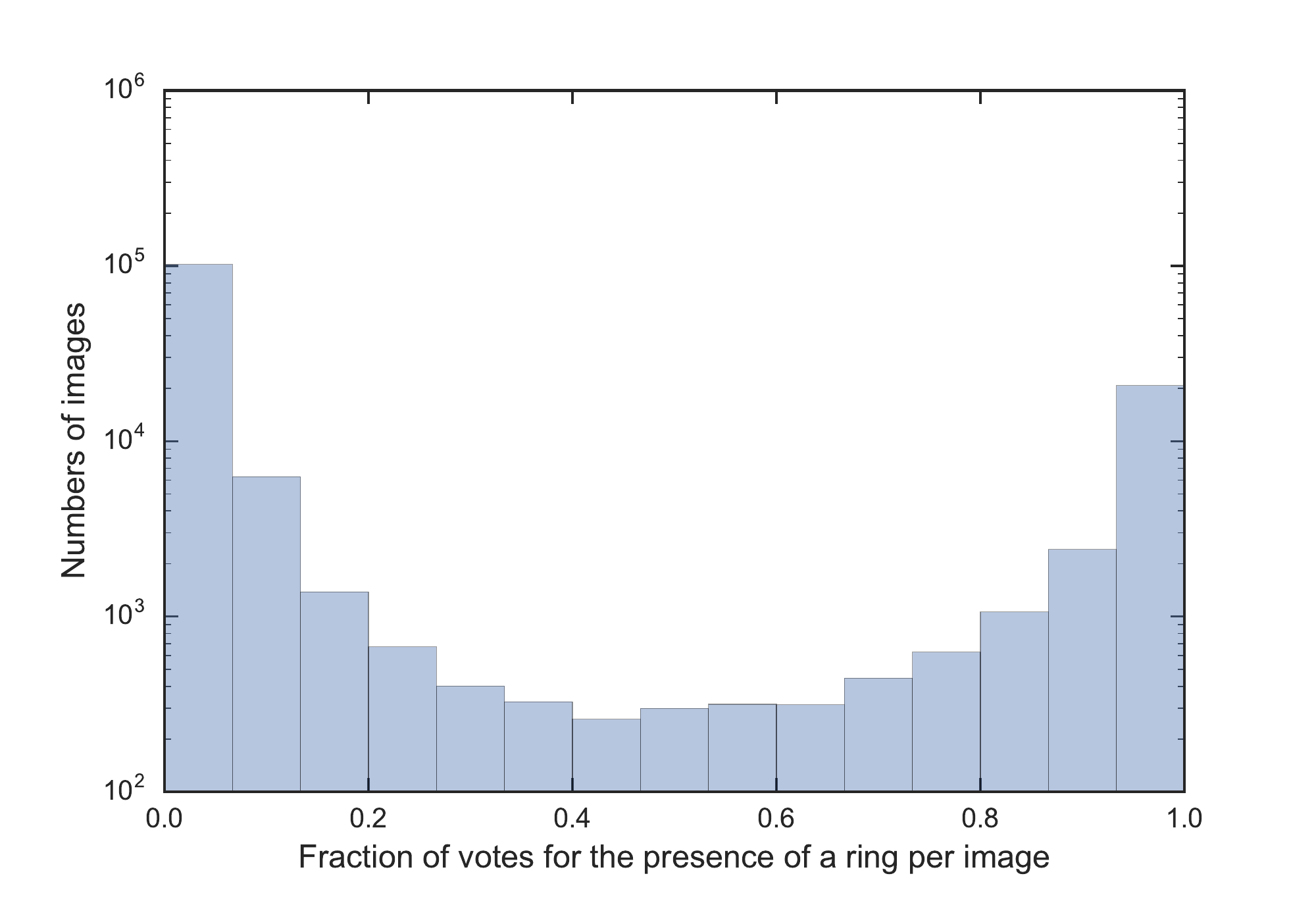}
\end{center}
\caption{\label{fig:Classifications}Left: the distribution of the number of classifications each user made. Right: distribution of votes for muon presence in the images.}
\end{figure}

Figure~\ref{fig:Classifications} summarizes the input we received from the volunteers of Muon Hunter. The distribution of the number of classifications each user made roughly follows a log-normal distribution (shown as the red dashed curve), with a median of 30 images per user. Assuming the number of classifications from one user is roughly proportional to the time spent, this log-normal distribution is of similar nature to the dwell time of internet users on social media articles \cite{MuonHunterICRC}. There are 16 volunteers who classified more than 10,000 images, while there are 724 volunteers who only classified one image. 
Of 134,000 images 12\% had unanimous votes that a ring was present, 73\% were unanimous a ring was not present, and the remainder were split. The 15 votes for each image allowed us to estimate the confidence of the users’ classifications. The distribution of the fraction of votes for the presence of a ring in each image is shown in the right plot of Figure~\ref{fig:Classifications}.  

The volunteers are also engaged through an active social media presence including an online blog\footnote{\texttt{muonhunterblog.wordpress.com} (to date 1301 reads from 55 countries)} giving details about the experiment from the researchers involved; Facebook\footnote{\texttt{www.facebook.com/muonhunters/ (to date 103 likes and 104 follows)}} and Twitter\footnote{\texttt{twitter.com/ZooniMuonHunter (to date 25 followers and 7 likes)}} feeds; and material like postcards for dispersal at public outreach events, e.g. school fairs, and the FLWO visitor center. The Muon Hunter Launch event was also advertised on Facebook for a week, the \$15 ad reached 9,299 people in 50 countries and received 80 event responses. The tracking aspects of these social media sites can also help us to get some insight into the demographics of the people we are reaching and engaging: for example as a function of gender and age group for our Facebook page, see Table~\ref{tab:Facebook}, we reached an audience composed close to parity of 47\% of women and 52\% men, but see that we engaged (with the Facebook page) only 35\% women and 64\% men. These insights can help us understand where we need to focus in the future to even out or encourage participation.

\begin{table}
\caption{\label{tab:Facebook}Muon Hunter Facebook page reach and engagement as a function of age group and gender.}
\begin{center}
\begin{tabular}{c|cc|cc}
\br
Age Group & \multicolumn{2}{c}{People Reached} & \multicolumn{2}{c}{People Engaged} \\
                  & Female & Male & Female & Male \\
\mr
13-17 & 20\% & 19\% & 23\% & 32\% \\
18-24 & 13\% & 17\% & 4\% & 14\% \\
25-34 & 7\% & 10\% & 5\% & 8\% \\
35-44 & 3\% & 3\% & 2\% & 5\% \\
45-54 & 2\% & 1\% & 0.629\% & 2\% \\
55-64 & 0.91\% & 0.836\% & 0.629\% & 2\% \\
65+ & 0.495\% & 0.579\% & 0\% & 0.629\% \\
\br
\end{tabular}
\end{center}
\end{table}

\section{CNN: training \& evaluation}
One purpose of this project is to train a reliable CNN model to classify muon rings. 
Two sources of labels, one provided by the VERITAS analysis \cite{StandardAnalysis} and the other by the Muon Hunter user input, can be used for the training, validation, and testing of the models. 
We randomly selected 16 observation runs, each 30 minutes in duration, and analysed them using one of the standard VERITAS data analysis packages, with the muon identification proceedure outlined in \cite{MuonHunterICRC}. A detailed description of the CNN model can be found in \cite{CNN}. 
Treating all images with 10 or more of the 15 votes for muons as muon events and the rest as non-muon events, we were able to train a CNN model from the Muon Hunter volunteers with a test accuracy of $\sim$97\%; comparing favourably to the VERITAS standard analysis labels test accuracy of $\sim$95\%. 

\section{Summary and future work}
Citizen science is a great resource for both outreach and practical science. Having a simple, clearly defined project task helped a lot with the success of the Muon Hunter project, for which we received a phenomenal response from our volunteers. The input from the volunteers not only helped us train a more efficient machine learning model and gain insight into where the standard analysis may be lacking, it also identified interesting new avenues to research on interesting and/or unexpected images that would be missed in a purely machine processed context which could, in turn, become the focus of future citizen science projects.

\ack
The authors gratefully acknowledge all the Muon Hunter volunteers who contributed to this effort without whom this work would not be possible. Muon Hunters was developed with the help of the ASTERICS Horizon2020 project. ASTERICS is a project supported by the European Commission Framework Programme Horizon 2020 Research and Innovation action under grant agreement \textnumero 653477. VERITAS research is supported by grants from the U.S. Department of Energy Office of Science, the U.S. National Science Foundation and the Smithsonian Institution, and by NSERC in Canada. We acknowledge the excellent work of the technical support staff at the Fred Lawrence Whipple Observatory and at the collaborating institutions in the construction and operation of the instrument. The VERITAS Collaboration is grateful to Trevor Weekes for his seminal contributions and leadership in the field of VHE gamma-ray astrophysics, which made this study possible.

\section*{References}
\bibliography{MuonHunter}

\end{document}